\documentclass[a4paper]{jpconf}
\usepackage{graphicx}
\usepackage{amsmath}
\usepackage{xcolor}
\renewcommand{\bi}{\begin{itemize}}
\newcommand{\ei}{\end{itemize}}
\newcommand{\GeV}{\mathrm{GeV}}
\newcommand{\MeV}{\mathrm{MeV}}
\newcommand{\fm}{\mathrm{fm}}
\begin{document}
\title{Exploring the tension between nature and the Standard Model: the muon g-2}

\author{M Krsti\'c Marinkovi\'c$^1$ and N Cardoso$^2$}
\address{$^1$ 
 Trinity College Dublin,
             School of Mathematics,
	    Dublin 2, Ireland
}
\address{$^2$ 
CeFEMA, Departamento de F\'isica, Instituto Superior T\'ecnico (Universidade Te\'ecnica de Lisboa), Av. Rovisco Pais, 1049-001 Lisboa, Portugal}

\ead{mmarina@maths.tcd.ie, nuno.cardoso@tecnico.ulisboa.pt}

\begin{abstract}
Anomalous magnetic moment of the muon (muon g-2) is one of the most precisely measured quantities in particle physics. At the same time, it can be evaluated in the Standard Model with an unprecedented accuracy. The Muon g-2 experiment at Fermilab has started the major data collection and the aimed four-fold increase in precision will shed light on the current discrepancy between the theory prediction and the measured value. This renders a comparable improvement of the precision in the SM theory an essential ingredient in order to fully exploit the expected increase of precision in experimental results. For all these reasons, the muon g-2 is considered to be a great testing ground for new physics. 

Hadronic contributions are the dominant sources of uncertainty in the theoretical prediction of the muon g-2. A reciprocal effort to a precise determination of the leading hadronic contribution to the muon g-2 using lattice gauge theories is a direct measurement of the hadronic contributions to the running of the fine structure constant recently proposed by the MUonE experiment. A hybrid strategy including both experimental and lattice data sets is expected to give an independent check of the dispersive results from e+e- annihilation, which dominate the current world average.
\end{abstract}

\section{Introduction}
\medskip
Muon magnetic moment represents one of the most sensitive tests of the Standard Model (SM), 
as this observable can be extremely accurately measured in the experiment 
and it is also predicted in SM with high precision. 
The deviation of muon's gyromagnetic ratio from 
the Dirac's relativistic quantum-mechanical prediction ($g_l^{(0)}=2$,  $l=e,\mu,\tau$)
is generally parametrized by { {\it the anomalous magnetic moment}},  $a_{\mu}=(g_{\mu}-2)/2$, also known as the {\it muon g-2} anomaly. 

These are particularly exciting times for the muon $g-2$! On the experimental side, the ongoing measurements of this quantity at the E989 experiment at Fermilab will achieve a fourfold improvement in precision in the next couple of years.   
A comparable measurement with different set of systematics is planned for the E34 experiment, for which the first tests are being performed at at J-PARC \cite{Abe:2019thb}. 
A careful reconsideration and further improvement 
of the 
SM
theory precision is essential in order to fully exploit the expected greater precision of experimental results and confirm or disprove the new physics indicated by the current more than 3--sigma discrepancy between the E821 experiment at BNL \cite{Bennett2006} and the theoretical estimate \cite{Keshavarzi:2018mgv, Davier2011, Jegerlehner:2018zrj}.

The dominating uncertanties in the SM prediction come from the hadronic-light-by-light contribution and the leading hadronic contribution: hadronic vacuum polarisation (HVP)  \cite{Keshavarzi:2018mgv}. 
The improved accuracy in the leading hadronic contribution to the muon $g-2$ can be achieved by combining experimental input from R-ratios with the lattice QCD data \cite{BerneckerMeyer, RBC2018}. 
A comparable precision of the result from first principles requires a continuation of the steady theoretical and algorithmic advances and lattice QCD community is investing significant efforts towards achieving this goal. For recent reviews see Refs.\,\cite{Meyer:2018til, Miura:2019xtd, GuelpersProc}.   
 
A reciprocal way to obtain the HVP is to directly measure the running of the fine structure constant 
in the space-like momenum region, as proposed in Ref.\,\cite{CarloniCalame2015}. The 
 idea was further developed into a novel proposal to measure HVP contribution to the muon g-2
from the space-like scattering of high-energy muons on the fixed electron target \cite{MUonEproposal}, which later evolved into a MUonE project.  The effort around the MUonE project initially became part of the Physics Beyond Colliders program at CERN \cite{QCDPBCreport}  and lead to the formation of the MUonE collaboration and submission of the Letter of Intent to the SPS and PS Committee at CERN \cite{CDSLoI}.  Lattice QCD and MUonE will measure the HVP contribution to the muon g-2 to high accuracy in complementary momenta ranges. Thus, a hybrid strategy including both  output of MUonE experiment and lattice data sets is expected to give an independent check of the current results based on dispersion approach, which rely on the input from $e^{+}e^{-}$ annihilation experiments.  

In Section\,\ref{sec:spacelike} of this proceedings contribution
we discuss the relation between the running of the QED coupling $ \alpha(Q^2)$ and the HVP contribution to the muon g-2, on which the idea of MUonE project is based. Section\,\ref{sec:MUonE} outlines 
the MUonE experimental setup and main challenges.  
In Section\,\ref{sec:hybrid} we propose a strategy to combine the MUonE and lattice results into a precise estimate of the HVP contribution to the muon g-2.  Section\,\ref{sec:calc} discusses the technicalities of our calculation. This is followed by the discussion of preliminary results  in Section\,\ref{sec:results}, and a summary in Section\,\ref{sec:summary}. 
\section{Space-like evaluation of the leading 
hadronic contribution to the muon $g-2$} 
\label{sec:spacelike}
\medskip

Current best estimates of the leading hadronic contribution to the muon $g-2$, $a_{\mu}^{HLO}$, are obtained by computing the dispersion integral of the hadron production cross sections in $e^{+}e^{-}$ annihilation (R-ratios) at low-energies  \cite{Keshavarzi:2018mgv, Davier2011, Jegerlehner:2018zrj}. 
The precision of the dispersive approach via time-like data  is thus governed by the experimental input and low energy effective modeling for a number of regions and resonances in R(s). The dependance of the leading hadronic contribution to the vacuum polarisation $\Pi(Q^2)$ on the space-like momenta  can also be related to  R(s) 
via the once subtracted dispersion relation
\cite{Blum2002, Lautrup1968}
\begin{align}
\Pi(Q^2)-\Pi(0)=\frac{Q^2}{3\pi} \int_{0}^{\infty} ds \frac{R(s)}{s(s-Q^2)},
\label{eq:subtractedpol}
\end{align}
which in turn gives the leading hadronic contribution to the muon $g-2$ 
\begin{align}
a_{\mu}^{HLO}= \frac{\alpha}{\pi}\int_{0}^{1}dx (1-x) [\Pi(Q^2(x))-\Pi(0)], ~~~~~~ \mathrm{where} ~~
Q^2(x)=\frac{x^2 m_{\mu}^2}{1-x} >0
\label{eq:spacelike}
\end{align}
is a space-like (Euclidean) momentum. 
Since the infrared subtracted vacuum polarisation defines the running of the effective electromagnetic coupling
\begin{align}
\alpha[Q^2(x)] = \frac{\alpha(0)}{1- \Delta \alpha[Q^2(x)]}, ~~~~~~~~ 
\Delta \alpha[Q^2(x)]= 4\pi^2 \mathrm{Re} [\Pi(Q^2(x))-\Pi(0)],
\end{align}
we obtain the relation between the leading hadronic contribution to the muon g-2 and the running of the effective QED coupling \cite{Lautrup1968}
\begin{align}
a_{\mu}^{HLO}= \frac{\alpha}{\pi}\int_{0}^{1}dx (1-x) \Delta \alpha_{had}[Q^2(x)].
\label{eq:spacecoupling}
\end{align}
Eq.\,\ref{eq:spacecoupling} and the value of  $a_{\mu}^{HLO}$ from the dispersive approach \cite{Keshavarzi:2018mgv} is used
to determine one of the most sought-after electroweak SM  parameters:  $\Delta \alpha_{had}(M_Z^2)$. And the opposite holds true, a precise knowledge of the running of the effective QED coupling $\Delta \alpha_{had}[Q^2(x)]$ at low momenta would in turn give a precise determination of the $a_{\mu}^{HLO}$. 
The  MUonE's strategy to perform such high-precision measurement of the total $\Delta \alpha(Q^2)$ is
outlined in the next section and discussed in more detail in 
Ref.\,\cite{CDSLoI}.
 The running contributions to $\Delta \alpha(Q^2)$ can be split into the hadronic and the leptonic part:
\begin{align}
\Delta \alpha(Q^2)=  \Delta \alpha_{had}(Q^2) + \Delta \alpha_{lep}(Q^2).
\end{align}
Subtracting the purely leptonic  part, which is perturbatively known up to three
 loops in QED \cite{Steinhauser}, one measures directly $\Delta \alpha_{had}(Q^2)$.  

\begin{figure}
\includegraphics[width=0.42\textwidth]{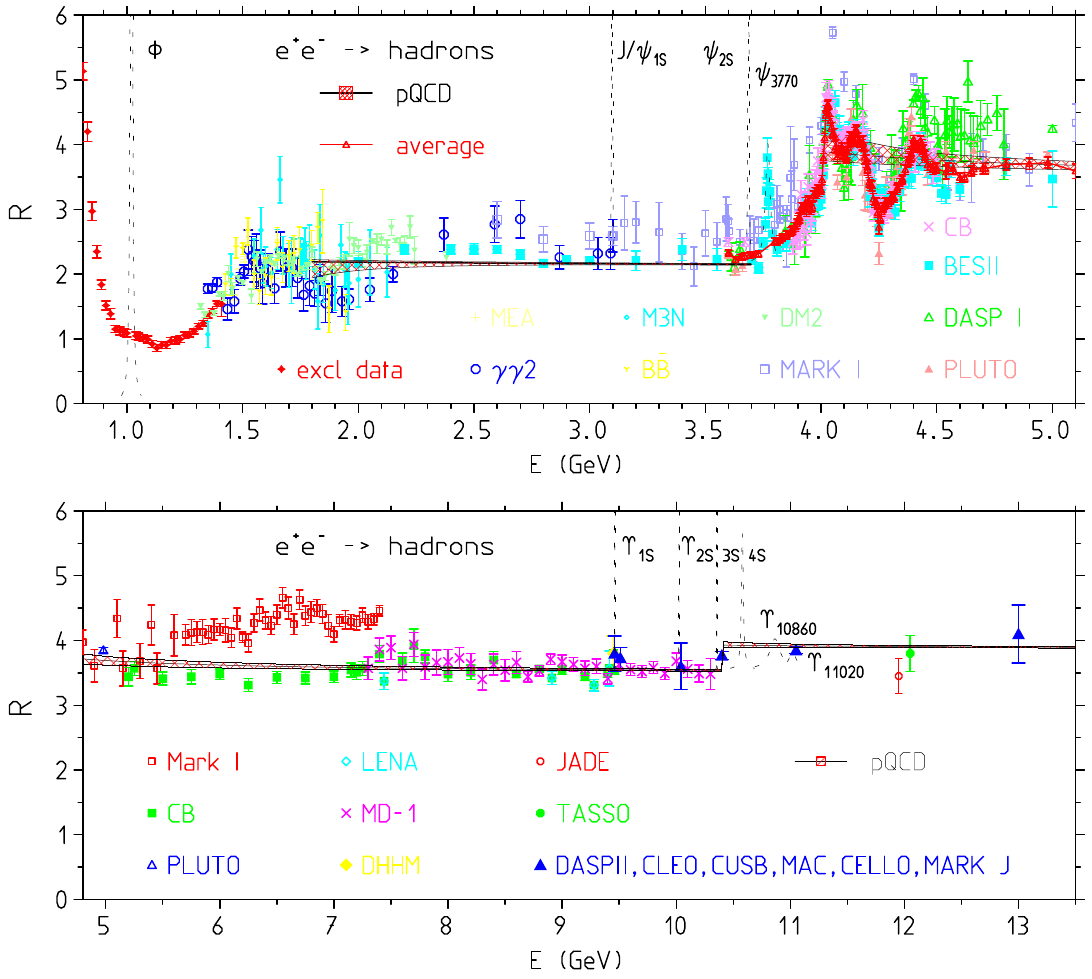}
\hspace*{0.01\textwidth}
\includegraphics[width=0.55\textwidth]{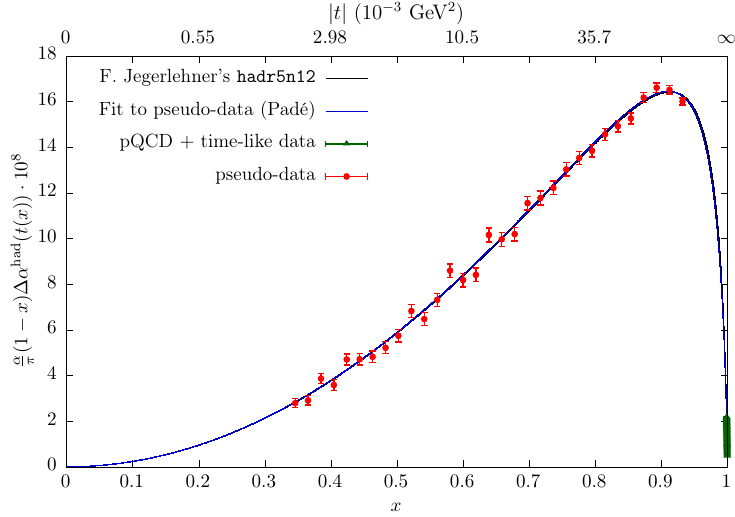}
\caption{Left panel:  R-ratio data from different $e^{+}e^{-}$ annihilation experiments  \cite{Jegerlehner:2009ry} that determine the leading hadronic contribution to the muon g-2 in the time-like regime (cf. 
Eq.\,\ref{eq:subtractedpol}). Right panel: Vacuum polarisation  integrand for the space-like momenta is a smooth function of $x$ and correspondingly $Q^2$. The pseudo data created utilising current uncertainties of R-ratios and the known input of the muon-electron cross sections gives an estimated statistical precision of the MUonE's measurement to be 0.3\% \cite{CarloniCalame2015}. The plot is reproduced from Ref.\,\cite{GrazianoICHEP}.}
\label{fig:timelike}
\end{figure}

While the time-like region shown in the left panel of Figure\,\ref{fig:timelike} is highly fluctuating due to resonances and  particle production threshold effects, the integrand in the space-like (Eq.\,\ref{eq:spacecoupling})  is a smooth and resonance-free function of the squared momentum transfer, 
as can be observed in the right panel of Fig.\,\ref{fig:timelike}. 
Exchanging the order of $x$ and $s$ integrations  in  Eq.\,\ref{eq:spacelike} leads to the form of the space-like integral commonly used for lattice QCD evaluations of the leading hadronic contribution to the muon g-2:
\begin{align}
a_{\mu}^{HLO}= \big (\frac{\alpha}{\pi}\big )^2  \int_{0}^{\infty} dQ^2 f(Q^2) \hat{\Pi}(Q^2),
\label{eq:HVPintegral}
\end{align}
where $f(Q^2)$ is a known, positive kernel function of the muon mass as originally introduced in 
Ref.\,\cite{Blum2002}. 
We will refer to the Eq.\,\ref{eq:HVPintegral} in the following text as the {\it HVP integral}. 

\section{Muon-electron elastic scattering: MUonE} 
\label{sec:MUonE}
\medskip

The MUonE experiment plans to perform a direct {\it space--like} measurement of the $\Delta \alpha_{had}(Q^2)$, and utilise a space-like integral from Eq.\,\ref{eq:spacecoupling} 
 to obtain the leading hadronic contribution to the muon g-2. 
 High-energy muons  will be scattering on atomic electrons of a low-Z target, such as Beryllium. 

A proposed scheme of MUonE experimental design can be found in Ref.\,\cite{CDSLoI}. Each module would consist of a Beryllium target and three tracking layers made of two silicon modules. Approximately 40 modules are needed in order to reach the goal precision in $a_{\mu}^{HLO}$. This would lead to the intrinsic angular resolution of $0.02~\mathrm{mrad}$. 
$\Delta \alpha (Q^2)$ in the space-like region can then be obtained from the differential cross section of the elastic process $\mu e \rightarrow \mu e$,  measured as a function of the squared momentum transfer. 
Further details on the progress of the detector design and anticipated systematics can be obtained from
Refs.\,\cite{QCDPBCreport, CDSLoI,GrazianoICHEP, Marconi:2019rio}. The results of the commissioning phase of the feasibility test at the CERN North Area (using COMPASS beamline) are reported in Ref.\,\cite{Ballerini:2019zkk}. It is important to note that in order to know $\Delta \alpha_{had}(Q^2)$ with the aimed precision of $10\mathrm{ppm}$, the ratios of the Standard Model cross-sections for the muon electron space-like scattering have to be known with a precision $O(10^{-5})$.
A great amount of theoretical efforts has already gone into computation of the cross section ratios
to such a high precision. To name a few recently completed studies dedicated to the muon-electron scattering process, the NNLO QED corrections have been  computed in Refs.\,\cite{Primo1, Primo2}, a subset of hadronic corrections at NNLO has been computed in Ref.\,\cite{Fael1} and Ref.\,\cite{{Pavia}} discusses the QED and purely week corrections at NLO.
\section{Hybrid approach: preliminaries}
\label{sec:hybrid}
\medskip
The MUonE experiment is  expected to give the value of $a_{\mu}^\mathrm{HLO}$ in the momenta range $[0, Q^2_{exp,max}]$, $Q^2_{exp,max}\approx 0.14\GeV^2$, with a statistical precision of roughly 0.3\% after two years of data taking. This corresponds to the range $[0,0.93]$ in variable $x$ defined in Eq.\,\ref{eq:spacelike}  and gives 
about 90\% of the total HVP integral. 
On the other hand, due to the finite volume momentum quantization, lattice QCD measures the HVP in a momentum region complementary to the range accessible in the MUonE experiment. Thus, a hybrid strategy including both experimental and lattice data sets is expected to give an alternative estimate of the HVP contribution obtained solely from the lattice or from e+e- annihilation. The proposed strategy is inspired by the approach described in Ref.\,\cite{Golterman}, which allows for a better control of the systematics in the HVP contribution by dividing the space-like $Q^2$-range into three sub-intervals and probing different integration boundaries of these domains. 
The maximum momentum directly reachable in the MUonE experiment suggests to break up the computation of the HVP integral into the computations of three sub-integrals
\begin{align}
a_{\mu}^\mathrm{HLO}&= I_0+I_1+I_2 , \label{eq:subintegrals}\\
I_0&=\frac{\alpha}{\pi}\int_{0}^{0.93...}dx (1-x) \Delta \alpha_{had}[Q^2(x)], \label{eq:I0}  \\ 
I_1&=\left(\frac{\alpha}{\pi}\right)^2 \int_{0.14}^{Q^2_{max}}dQ^{2} f(Q^2) \times \hat{\Pi}(Q^2),\label{eq:I1} \\
I_2&=\left(\frac{\alpha}{\pi}\right)^2 \int_{Q^2_{max}}^{\infty}dQ^{2} f(Q^2) \times \hat{\Pi}_{pert.}(Q^2),\label{eq:I2} 
\end{align}
where $\hat{\Pi}(Q^2)=\Pi(Q^2)-\Pi(0)$ denotes the infrared subtracted vacuum polarisation and 
$Q^{2}_{max}$ is the momentum value above which the perturbation theory is applied. 
Due to a highly peaked integrand in the low-$Q^2$, the data corresponding to the integration range of $I_0$ 
are crucial for the overall precision of the HVP space-like integral. 
However, sub-percent determination of the total HVP requires the sum of the  integrals $I_1$ and $I_2$ to be determined with a relative precision $O(10^{-1})$ or higher. 
 \begin{figure}[t!]
\centering
\includegraphics[width=0.55\textwidth]{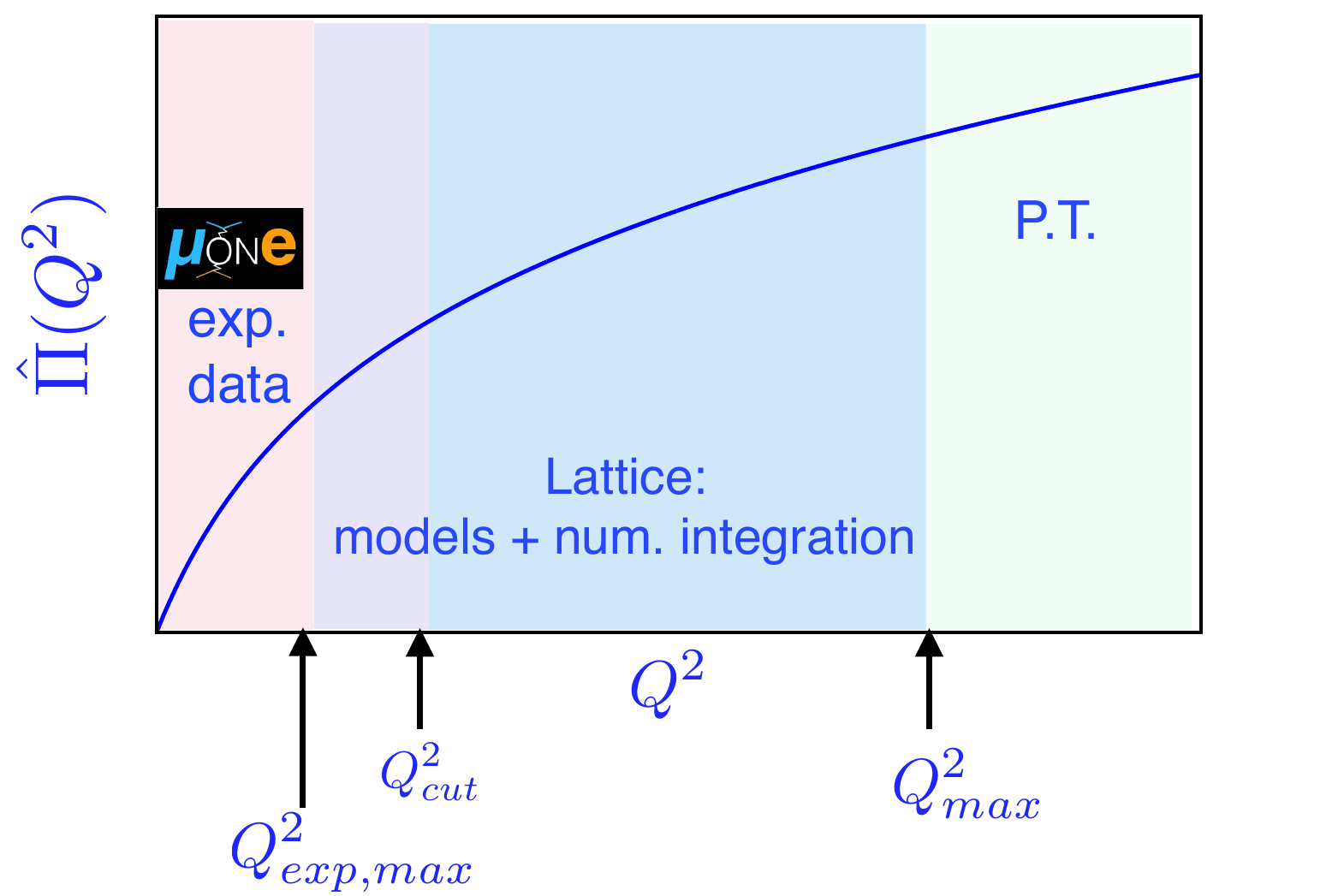}
\caption{A sketch of the hybrid approach, which would combine MUonE data in the low-$Q^2$ range: $[0,Q^2_{exp,max}]$ with the lattice data in the intermediate-$Q^2$: $[Q^2_{exp,max},Q^2_{max} ]$ and match to perturbation theory at some value $Q^2_{max}$. The intermediate momentum range is additionally divided into two complementary ranges: $[Q^2_{exp,max},Q^2_{cut} ]$  and $[Q^2_{cut},Q^2_{max} ]$, where $Q_{cut}$ is varied to obtain a reliable estimate of the systematics, as suggested in Ref.\,\cite{Golterman}.}
\label{fig:Hybrid}
\end{figure}
\section{Calculation details}
\label{sec:calc}
\medskip
As a proof of concept, we perform a computation of the HVP integral reduced to the intermediate momentum range ($I_1$, as defined in Eq.\,\ref{eq:I1}) on a set of six $N_f=2$ CLS gauge ensembles with pion masses from $185\,\MeV$ to $437\,\MeV$. 
The generation of gauge configurations featuring O(a) improved Wilson fermions and Wilson gauge action was performed using DD-HMC and MP-HMC algorithms. The measurements of the hadronic vacuum polarisation are performed with a modified version of the DD-SCOR measurement suite. The measured gauge ensembles span a range of lattice spacings from $0.05\,\fm$ to  $0.08\,\fm$, which allows for a safe continuum extrapolation. The  unphysical pion masses of the chosen gauge ensembles 
necessitate chiral extrapolations to the physical value of $m_{\pi}$,  as will be discussed in the following. 

As we are working with $N_f=2$ ensembles, strange and charm quark vacuum polarisation are partially quenched. The values of $\kappa_s$ and $\kappa_c$ used in the measurements 
were not specifically tuned for this study, but rather taken from Ref.\,\cite{Mainz2017}.
In this first estimate of $I_1$, our calculation neglects isospin breaking effects and the disconnected contribution, as they are expected to scale and enter at a sub-permille precision. However, the technology to compute these corrections has been developed and we plan to include it in forthcoming studies. 

\section{Results and discussion}
\label{sec:results}
\medskip
We choose to use off-diagonal components of the  vacuum polarisation tensor, in order to obtain the scalar vacuum polarisation function
\begin{align}
{\Pi}(Q^2)=\frac{{\Pi}_{\mu\nu}(Q)}{q_{\mu}q_{\nu}},\quad\quad\quad\quad \mu\neq\nu, 
\label{eq:unsub}
\end{align}
for non-zero values of momentum squared.
 \begin{figure}[t!]
\centering
\includegraphics[width=1.00\textwidth]{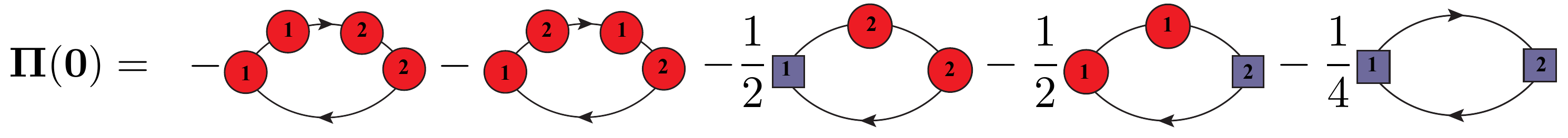}
\caption{Diagrammatic representation of the expansion of a vector-vector correlator around $Q^2=0$, which is used to compute $\Pi(0)$. Red circles represent the insertion of the point-split vector current and gray squares represent the insertion of the tadpole current operator. This approach has been first proposed in Ref.\,\cite{Divitiis}.}
\label{fig:Vacpol_zero}
\end{figure}
 \begin{figure}[b!]
\centering
\includegraphics[width=1.00\textwidth]{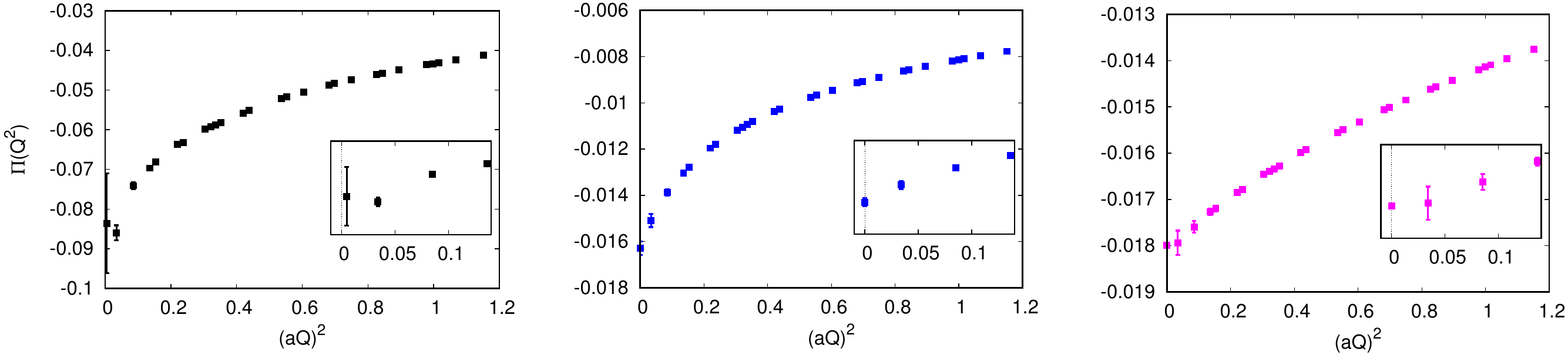}
\caption{Unsubtracted hadronic vacuum polarisation for light (left panel), strange (middle panel) and charm (rignt panel) contributions computed using the expression in Eq.\,\ref{eq:unsub}. The  lattice size is $48^3 \times 96$, lattice spacing $a=0.066\,\fm$ and the corresponding pion mass is $m_{\pi}=311\,\MeV$.}
\label{fig:Vacpol_unsub}
\end{figure}
This allowed a test of the method proposed  in Ref.\,\cite{Divitiis} on a wide range of gauge ensembles, 
where the vacuum polarisation at zero momentum  is obtained by computing a mixed derivative of the electromagnetic current two--point correlation function 
\begin{align}
 \Pi(0)=-\frac{\partial^2 \Pi_{\mu \nu}(Q)}{\partial q_{\mu} \partial q_{\nu}}|_{Q^2=0}, 
\label{eq:unsub}
\end{align}
which corresponds to the sum of operator insertions depicted in Figure\,\ref{fig:Vacpol_zero}. 
A first look at the data indicates that that the method exibits sufficient precision at zero momenta for strange and charm contribution, as demonstrated in Figure\,\ref{fig:Vacpol_unsub}. On the other hand, the tested method did not prove  to be effective for the light quark contribution.

Combined continuum and chiral extrapolation of the individual light\,(l), strange\,(s) and charm\,(c) quark contribution to the integral  over the intermediate momenta in the region $[0.14,4]\,\GeV^2$ on the selected set of ensembles gives 
\begin{align} 
\sum_i I^{i,cont}_1=79.5(3.5) \cdot 10^{10}, ~~~~~~ i=\mathrm{l},\mathrm{s},\mathrm{c}.
\end{align}
The obtained $\approx 4.4\%$ uncertainty includes the error of continuum and chiral extrapolation, as well as  systematics due to variation of $Q^2_{cut}$. Note that this preliminary absolute error in $I_1$  corresponds to $0.5\%$ of the full HVP contribution, which is roughly the aimed precision of the MUonE experiment. 
Due to the lack of $O(a)$ improvement in the vector current, the continuum extrapolation is linear in the lattice spacing
\begin{align}
I_1^{j}&=\alpha_1+\alpha_2\,m_{\pi}^2 + 
\alpha_3\,m_{\pi}^2\, ln(m_{\pi}^2)+
\alpha_4\,a,~~~~~~ j=\mathrm{l},\mathrm{s};\\
I_1^\mathrm{c}&=\beta_1+\beta_2\,a,
\end{align}
 despite the fact that the QCD action is O(a) improved. In order to reduce this source of error, improvements along the lines of \cite{Gerardin:2018kpy} are advisable. 
\section{Summary and outlook}
\label{sec:summary}
\medskip
This writeup discusses how the novel proposal to determine the leading hadronic contribution to the muon g-2 in MUonE experiment \cite{CarloniCalame2015,CDSLoI} can be combined with the lattice determination of HVP in the intermediate momentum region, which is not directly accessible in the MUonE experimental setup. Similar determinations have been reported in Refs.\,\cite{Giusti1,Giusti2} using time momentum representation and twisted-mass discretisation of the fermion QCD action. 
In Ref.\,\cite{Giusti2}, the authors report the sum of $I_1$ and $I_2$ with an impressive precision of $2\%$ and taking into account the isospin breaking effects. 

The preliminary result of the intermediate integral $I_1$ reported here gives a first estimate on $N_f=2$ ensembles using O(a) improved Wilson fermions. 
While chiral and continuum extrapolation of the connected contribution to the integral $I_1$ have been performed, the effects of isospin breaking and the disconnected contribution have been ignored so far, and will be included in consecutive works. Our determination of $I_1$ lacks the estimate of the finite volume error, which might be substantial. These effects in the framework of purely lattice QCD calculations of the HVP were discussed in Refs.\,\cite{Aubin:2015rzx,PACS,Hansen:2019rbh, Mainz2019, Aubin:2019kiq}. Furthermore, in this work the contribution from the perturbative $Q^2$ region ($I_2$) has not been estimated and $Q^2_{max}$ has been set to $4\GeV^2$. The estimate of $I_1+I_2$ with full systematics would require an additional variation of $Q^2_{max}$. It is well known  that the $I_2$ contribution is much smaller than 1\% of the total HVP  \cite{ Mainz2017,Mainz2019,Borsanyi:2017zdw}, due to a high supression of the high-momenta contribution by the integration kernel $f(Q^2)$. 

The proposed hybrid approach would complement the effort of lattice  gauge theories  \cite{Meyer:2018til, Miura:2019xtd, GuelpersProc} to compute HVP integral from first principles. 
If the MUonE proposal is successful, a pilot run planned in the time frame of two years aims at validating the outlined ideas and the full detector construction is expected by the end of 2022 \cite{CDSLoI}, which will be followed by a two year period of data taking. The obtained preliminary result demonstrates that  full control over the systematics in the lattice computation of contribution $I_1$ is attainable in this time frame. 
The expected high-accuracy of MUonE experiment in the low$-Q^2$ region, combined with the lattice input in the intermediate$-Q^2$ would then lead to a precise estimate of the HVP entirely independent of the dispersive approach and the input of the $e^+e^-$ annihilation experiments. 
\section{Acknowledgements}
\medskip
It is a pleasure to thank the organisers for the invitation to a very interesting conference and for their great organisation of the event. M.K.M. gratefully acknowledges help in preparing the slides on the MUonE project by Carlo Carloni Calame, Massimo Passera, Fulvio  Piccinini and Graziano Venanzoni. 
Many thanks to  
Pedro Bicudo
 for critical reading of the manuscript.
We are grateful to the CLS initiative for access to gauge ensembles and to Nazario Tantalo for sharing the first version of DD-SCOR suite that includes vacuum polarisation measurements. This work is supported by Research IT, CERN IT, and Altamira. N.C. is supported by FCT grant~{SFRH/BPD/109443/2015} and acknowledges the support of
CeFEMA under the FCT contract for R\&D Units
UID/CTM/04540/2013 and the FCT project grant
CERN/FIS-COM/0029/2017.

\end{document}